\documentclass[prb,superscriptaddress,twocolumn,amsmath,amssymb,showpacs]{revtex4}
\usepackage{graphicx}

\begin{document}
\title{Mechanism for puddle formation in graphene} 

\author{S. Adam}
\affiliation{Center for Nanoscale Science and Technology, National Institute of Standards and Technology, Gaithersburg, MD 20899, USA}
\author{Suyong Jung}
\affiliation{Center for Nanoscale Science and Technology, National Institute of Standards and Technology, Gaithersburg, MD 20899, USA}
\affiliation{Maryland NanoCenter, University of Maryland, College Park, MD 20472, USA} 
\author{Nikolai N. Klimov}
\affiliation{Center for Nanoscale Science and Technology, National Institute of Standards and Technology, Gaithersburg, MD 20899, USA}
\affiliation{Maryland NanoCenter, University of Maryland, College Park, MD 20472, USA}
\affiliation{Physical Measurement Laboratory, National Institute of Standards and Technology, Gaithersburg, MD 20899, USA}
\author{Nikolai B. Zhitenev}
\affiliation{Center for Nanoscale Science and Technology, National Institute of Standards and Technology, Gaithersburg, MD 20899, USA}
\author{Joseph A. Stroscio}
\affiliation{Center for Nanoscale Science and Technology, National Institute of Standards and Technology, Gaithersburg, MD 20899, USA}
\author{M. D. Stiles}
\affiliation{Center for Nanoscale Science and Technology, National Institute of Standards and Technology, Gaithersburg, MD 20899, USA}

\date{\today}


\begin{abstract}  
When graphene is close to charge neutrality, its energy landscape is
highly inhomogeneous, forming a sea of electron-like and hole-like
puddles, which determine the properties of graphene at low carrier
density.  However, the details of the puddle formation have remained
elusive.  We demonstrate numerically that in sharp contrast to
monolayer graphene, the normalized autocorrelation function for the
puddle landscape in bilayer graphene depends only on the distance
between the graphene and the source of the long-ranged impurity
potential.  By comparing with available experimental data, we find
quantitative evidence for the implied differences in scanning tunneling microscopy
 measurements of electron and hole puddles for monolayer and bilayer
graphene in nominally the same disorder potential.
\end{abstract}
\maketitle
\normalsize

\section{Introduction}

In monolayer graphene, the hexagonal arrangement of carbon atoms
dictates that in the absence of atomic-scale disorder, graphene is a
gapless semiconductor\cite{kn:dassarma2011a,kn:neto2009} 
that is always metallic at low temperature.\cite{kn:fuhrer2009b}
This metallic behavior holds even in the presence of quantum
interference and strong disorder,\cite{kn:bardarson2007} in 
stark contrast to most other materials, which undergo a
metal-to-insulator transition at low carrier
density.\cite{kn:anderson1958,kn:tanatar1989}  The
physical origin
for this robust metallic state is that the ground-state of graphene at
vanishing mean carrier density becomes spatially inhomogeneous,
breaking up into electron-rich and hole-rich metallic regions
connected by highly conducting p-n junctions.\cite{kn:katsnelson2006b}
Bilayer graphene comprising two sheets of graphene 
that become strongly coupled due to the $AB$ 
stacking arrangement\cite{kn:mccann2006b}
shares some properties with regular semiconductors (such as the
parabolic band dispersion) and in other ways behaves like monolayer
graphene, including having chiral wavefunctions and forming
electron and hole puddles at low density.     

These electron and hole puddles have now been observed in several
experiments of exfoliated graphene on an insulating SiO$_2$ substrate including
Refs.~\onlinecite{kn:martin2008,kn:zhang2009,kn:deshpande2009b,kn:deshpande2011,kn:jung2010,kn:rutter2011}.
While these authors suggest that long-range charged
impurities in the substrate could be responsible for the spatial
inhomogeneity, detailed comparisons to microscopic models have not
been made. 

In this paper, we demonstrate that differences between the spatial
properties of puddles in monolayer and bilayer graphene can be
quantitatively explained by the differences in the screening
properties of the two systems (that ultimately arises from the
differences in their band-structure).  Numerical results show that the
correlation length for bilayer graphene is relatively independent of
density and significantly smaller than that of monolayer graphene for
a typical range of impurity densities.  Finally, we find good
quantitative agreement when comparing our results with available
experimental data.

The rest of the paper is organized as follows.  In
Sec.~\ref{Sec:Formalism}, we outline the the theoretical model,
providing a heuristic understanding of our results using the
Thomas-Fermi (TF) screening theory.  However, the TF significantly
underestimates the effect of electronic screening in both monolayer
and bilayer graphene. It is therefore necessary to use the Random
Phase Approximation (RPA) screening theory, which we discuss in
Sec.~\ref{Sec:RPA}.  Our main finding is that the puddle correlation
length in bilayer graphene $(\xi \approx 3.5~{\rm nm})$ is relatively
insensitive to the impurity concentration and carrier doping.  This is
in contrast to monolayer graphene, where the puddle correlation length
varies from $3~{\rm nm}$ in dirty samples to more than $35~{\rm nm}$
in clean samples.     

The comparison with experiment is done in
Sec.~\ref{Sec:Compare}, where we examine three different experimental results: (1) We
consider first the experimentally determined normalized correlation
function $A(r)$ (see definition below) obtained from the 
scanning tunneling microscopy (STM) data reported
for exfoliated bilayer graphene in Ref.~\onlinecite{kn:rutter2011}.
The full functional form of $A(r)$ agrees with the theory where the
only adjusted parameter in the theory is the distance $d$ of the impurities
from the graphene sheet.  In particular the experimentally determined
correlation length $\xi = (3.68 \pm 0.03)~{\rm nm}$, defined here as the
half-width at half-maximum (HWHM) decay length of $A(r)$ agrees well
with the $d=1~{\rm nm}$ RPA theory value of $\xi = 3.4~{\rm nm}$.  This
value of $d$ is both reasonable and consistent with those determined from other
transport measurements on bilayer graphene.\cite{kn:adam2008a}  (2)
From the monolayer graphene STM experimental data reported in
Ref.~\onlinecite{kn:jung2010}, we extract a correlation length $6~{\rm nm} < \xi < 11~{\rm nm}$.  Since the
measurements~\cite{kn:jung2010,kn:rutter2011} were made on the same
exfoliated graphene sample containing both single layer and bilayer
graphene regions, we expect that the extrinsic disorder potential is
statistically identical for the two samples.  Therefore, using the
value of $d=1~{\rm nm}$ (discussed above) and the disorder induced
Dirac point shift reported in Ref.~\onlinecite{kn:rutter2011}, we
calculate theoretically (without any adjustable parameters) that a
monolayer graphene sample in the same disorder environment would have
a puddle correlation length $\xi = 8~{\rm nm}$, in reasonable
agreement with the experiment.  (3) We then compare $A(r)$ obtained
using scanning Coulomb blockade spectroscopy reported in Ref.~\onlinecite{kn:deshpande2011} 
with our theoretical results
for monolayer graphene at the Dirac point.  The parameters used in the
theory were obtained from separate transport measurements on the same
experimental sample.\cite{kn:deshpande2011}  The agreement between
theory and experiment is remarkable since it involves no adjustable parameters.  Finally 
in Sec.~\ref{Sec:Conclude}, we conclude by making predictions for future experiments involving monolayer and
bilayer graphene on BN substrates.

\section{Formalism}
\label{Sec:Formalism}

The doping level of graphene can be measured in a variety of 
different ways.  In transport measurements, the gate voltage potential
that yields the resistivity maximum identifies the extrinsic doping
level due to extraneous sources, such as charged impurities in the
substrate impurities with density, $n_{imp}$.  While the width of the
resistivity maximum is a measure of the homogeneity of the
sample,~\cite{kn:adam2007a} or the electron-hole puddle distribution.

In local probe measurements, such as STM, the Dirac point energy
relative to the Fermi-level can be observed as a minimum in the
tunneling differential conductance, $dI/dV$ as a function of tunneling
bias.  Knowing the electronic dispersion relation (see details below),
for a particular gate voltage $V_g$, this spatial map $V(r)$ of the
Dirac point variation can then be used to extract the spatial
distribution of the local carrier density (characterized by a width
$n_{\rm rms}$).  We can also characterize the puddles through the radially 
averaged autocorrelation function
\begin{equation}
  C(r) = \frac{1}{2\pi}\int\limits_{0}^{2\pi} d\phi \langle\langle
  V({\bf r}) V(0) \rangle\rangle,
\label{Eq:auto}
\end{equation}
where the angular brackets denote an average over the image area and
the $\phi$-integration averages over orientations.  

Notice that while $C(0) = V_{\rm rms}^2$ (which is related to $n_{\rm rms}$)
characterizes the fluctuations in the puddle depth, $C(r)$ describes
the spatial profile of the electron and hole puddles.  We find it
useful to consider the normalized correlation function $A(r) =
C(r)/C(0)$.  We will argue below that $A(r)$ and $C(0)$ are quite
different physical quantities that depends quite differently on the
parameters of the extrinsic impurity potential.  (In addition, for a
typical STM experiment, where the shifts in the Dirac point are
determined~\cite{kn:zhang2009} from the shifts in $dI/dV$ at fixed
$V_g$, the determination of $C(0)$ is complicated by the experimental
uncertainty in converting spatial maps of $dI/dV$ to Dirac point energy shifts
$V({\bf r})$.  By contrast, the much smaller uncertainty in $A(r)$ is mostly
determined by the spatial resolution and image area.)  

To theoretically compute the correlation functions for puddles in
graphene, we make two assumptions.  First, the impurity potential
comes from a random two dimensional distribution of charged impurities
displaced by a distance $d$ from the plane with density $n_{\rm imp}$.  This 
model has been highly successful in describing the effect of
disorder in semiconductor heterojunctions\cite{kn:ando1982} and in 
graphene.\cite{kn:dassarma2011a}  This spatially varying potential gives rise to a varying charge
density and local variations in the screening of the potential.  Second,
we assume that it is possible to find a global screening
function $\epsilon(q, n_{\rm eff})$ that adequately describes the
effects of these local screening variations.  Here, the screening
depends on the disorder potential only through an effective carrier
density $n_{\rm eff}$.  This self-consistent screening model has been 
used previously to understand the minimum conductivity problem in both 
monolayer\cite{kn:adam2007a} and bilayer graphene.\cite{kn:adam2008a}  If these two assumptions
are satisfied, the correlation function, aside from a prefactor of
$n_{\rm imp}$, then depends only on the screened impurity potential
\begin{eqnarray}
\label{Eq:main}
C(r) &=&  2 \pi n_{\rm imp} \left( \frac{e^2}{\kappa} \right)^2 \int_0^\infty dq \frac{q \exp(- 2 q d)}{[q \epsilon(q, n_{\rm eff})]^2} J_0 (q r), 
\end{eqnarray}
where $\kappa$ is the bulk (3D) dielectric constant, $\epsilon(q)$ is
the surface (2D) screening function in the plane, $-e$ the electron
charge, and $J_0(x)$ is a Bessel function.  

\begin{figure}
\includegraphics[width=3.2in]{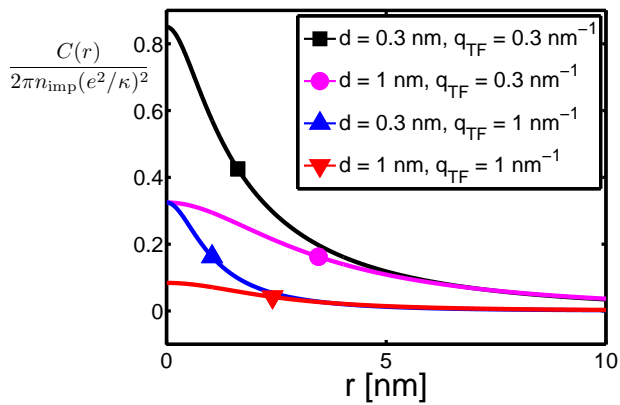}
\caption{\label{Fig:ThomasFermi} 
Theoretical calculations for the correlation function $C(r)$ using the
Thomas-Fermi approximation.  This autocorrelation function depends separately 
on the typical distance $d$ of long-ranged impurities from the graphene sheet, 
and $q_{\rm TF}$ the inverse effective screening length, allowing them to 
be determined independently.  Symbols show the correlation length $\xi$, defined as the HWHM length.}
\end{figure}

As an illustration, consider the Thomas-Fermi (TF) screening for which
the surface dielectric function is given by $\epsilon(q, n_{\rm eff})
= 1 + q_{\rm TF}(n_{\rm eff})/q$, where $q_{\rm TF}(n_{\rm eff})$ (discussed below) 
is the Thomas-Fermi screening wavevector.  Shown in Fig.~\ref{Fig:ThomasFermi}
is a calculation of $C(r)$ for different values of $q_{\rm TF}$ and
$d$.  Notice that for fixed $q_{\rm TF}$, the spatial dependence of
$C(r)$ depends on $d$ and $q_{\rm TF}$, but not on $n_{\rm imp}$.  On
the other-hand, the function $C(0)$ (which, within the TF can be
calculated analytically) depends on $n_{\rm imp}$,
$\kappa$, $d$ and $q_{\rm TF}$.  This is why we find it useful to use
the normalized correlation function $A(r)=C(r)/C(0)$ that describes
the spatial profile of the screened impurity potential.  
We also emphasize that
$A(r)$ contains different information than the typical puddle size --
for example, the puddle correlation length $\xi$ (recall 
that $\xi$ is defined as the HWHM of $A(r)$) describes the width of the screened impurity
potential, and not the mean impurity separation.  For example, in the low impurity
density limit, spatial maps of the puddles would show isolated
impurities, but $A(r)$ would not change (for fixed $q_{\rm TF}$).
Moreover, since $A(r)$ scales differently with $q_{\rm TF}$ and $d$,
in principle, both of these length scales can be extracted from a
measurement of $A(r)$.

Within the TF screening theory, any differences between monolayer and
bilayer graphene can only arise from differences in $q_{\rm TF}(n_{\rm
  eff})$.  The linear dispersion in monolayer graphene and the
hyperbolic dispersion in bilayer graphene gives rise to these
differences.  The low energy linearly dispersing bands of monolayer
graphene can be modeled by a single parameter, the Fermi velocity
$v_{\rm F}$, or equivalently, the effective fine-structure constant $r_s = e^2/(\kappa
\hbar v_{\rm F}) \approx 0.8$. $r_s$ characterizes the strength of the
electron-electron interaction for graphene on a SiO$_2$
substrate\cite{kn:dassarma2011a} and is useful because we are interested in 
the screening properties of graphene.  The Thomas-Fermi screening
wavevector is related to the density of states, and for monolayer
graphene is given by $q_{\rm TF}(n_{\rm eff}) = 4 r_s \sqrt{\pi n_{\rm eff}}$, where $n_{\rm eff}$
is the effective carrier density.

Bilayer graphene can be modeled with a hyperbolic dispersion with two
parameters $v_{\rm F}$ (throughout this manuscript, $v_{\rm F}$ is the
Fermi velocity of a single decoupled graphene sheet), and the
low-energy effective mass $m_{\rm eff}$.  For 
simplicity we use for the two parameters $r_s$ (defined above) 
and $n_0 = m_{\rm eff}^2 v_{\rm F}^2/(\hbar^2 \pi) \approx 2.3 \times 10^{12}~{\rm cm}^{-2}$ which 
is the characteristic density scale for the crossover from a (low density) parabolic 
to a (high density) linear dispersion.  The Thomas-Fermi screening wavevector for bilayer graphene
is given by $q_{\rm TF}(n_{\rm eff}) = 4 r_s \sqrt{\pi n_0} \sqrt{1 + n_{\rm eff}/n_0}$.      

\begin{figure}[!h]
\includegraphics[width=3.2in]{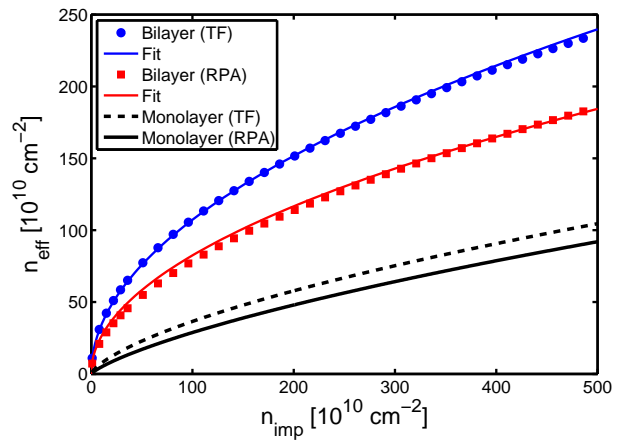}
\caption{\label{Fig:nrms} 
Effective carrier density as a function of impurity density assuming 
$d=1~{\rm nm}$, $r_s = 0.8$, and $n_0 = 2.3
\times 10^{12}~{\rm cm}^{-2}$.  For bilayer graphene, 
the blue circles show the Thomas-Fermi
approximation and the red squares are RPA results.  The empirical 
relation $n_{\rm eff} = \sqrt{n_{\rm imp}~n_1}$ adequately captures the RPA results,
with $n_1 = 6.8 \times 10^{11}~{\rm cm}^{-2}$.}
\end{figure}

As the system approaches the Dirac point, the fluctuations in carrier density
become larger than the average density.  In this case, screening
varies spatially with the density fluctuations.  We assume that it is possible to 
describe the effect of this screening by using the screening 
for an ideal system and an effective carrier density $n_{\rm eff}$ obtained
self-consistently.\cite{kn:adam2007a}  This is done by equating
the squared Fermi level shift with respect to the Dirac point with the
square of the potential fluctuations, $E^2[n=n_{\rm eff}] = C(0)$ 
where $C(0)$ is defined in Eq.~\ref{Eq:main} and $E[n] = \hbar v_{\rm F}
\sqrt{\pi n}$ for monolayer graphene, and $E[n] = v_{\rm F}^2 m_{\rm
  eff} \left[ \sqrt{1 + n/n_0} - 1 \right]$ for bilayer graphene.  The result
of this procedure are shown in Fig.~\ref{Fig:nrms}.

Within the TF theory, we can now qualitatively discuss the main
differences between monolayer and bilayer graphene.  For bilayer
graphene, the inverse screening length changes only slightly from the
low density value of $q_{\rm BLG} \approx 4 r_s \sqrt{\pi n_0}$ that
is set entirely by the band parameters.  As a consequence, the 
puddle correlation length {\it does not} change with the impurity concentration or carrier
density, and depends only on the distance $d$ of the bilayer graphene
sheet from the source of the long-ranged impurity potential.  In
contrast, for monolayer graphene, $q_{\rm MLG} = 4 r_s \sqrt{\pi
  n_{\rm eff}}$ depends essentially on $C(0)$ (and therefore on
$n_{\rm imp}$).  This heuristic description (which we make more quantitative
below) implies that depending on the sample quality, choice of
substrate, or doping, the puddle correlation length in monolayer graphene (but not
bilayer graphene) could vary by more than an order of magnitude.

\section{Random Phase Approximation}
\label{Sec:RPA}

While the TF screening theory discussed in the previous section is useful to obtain a
qualitative picture, we find that it significantly underestimates the
effect of electronic screening.  In both monolayer and bilayer graphene
it gives larger values for $C(0)$ and smaller values for $\xi$.  In what
follows we use the the random phase approximation (RPA) 
where the screening function is obtained using 
$\epsilon(q) = 1 + q_{\rm TF} {\tilde \Pi}(q)/q$.  

The normalized polarizability ${\tilde \Pi}(q)$ for monolayer\cite{kn:hwang2006b}
and bilayer\cite{kn:gamayun2011} graphene are both available in the 
literature.  We note that for monolayer graphene ${\tilde \Pi}(q)$ depends only on the dimensionless
variable $x = q/(2 \sqrt{\pi n_{\rm eff}})$
\begin{eqnarray}
{\tilde \Pi}(x) &=& 1 + \theta(x-1) \left[ \frac{\pi x}{4} - \frac{x}{2} {\rm arccsc}(x) -
 \frac{\sqrt{1 - x^{-2}}}{2} \right], \nonumber \\
 &\approx&  \theta(1-x) + \theta(x-1) \frac{ \pi x}{4}, 
\end{eqnarray} 
where $\theta(x)$ is a step-function and $D_0$ is the density of
states with $q_{\rm TF} = 2 \pi (e^2/\kappa) D_0 = 4 \sqrt{\pi n_{\rm
    eff}} r_s$.

In contrast, for bilayer graphene, the polarizability depends both on
the scaled momentum transfer $x$, and on $\eta = n_{\rm eff}/n_0 <8$,
which parameterizes the bilayer hyperbolic dispersion relation.  For
$\eta \ll 1$, the bilayer graphene dispersion is quadratic, while for
$1 \ll \eta \leq 8$, the dispersion is linear.  For $\eta > 8$, one
must consider the effects of a second higher-energy band that provides
additional screening\cite{kn:min2011a} and is not considered here.  By 
restricting the density to $n \leq 8 n_0$, we can simplify the expression 
for the bilayer polarizability reported in Ref.~\onlinecite{kn:gamayun2011}.
Using the bilayer density of states $D = D_0 \sqrt{1 + \eta}$, with $D_0
= 2 m /(\pi \hbar^2)$, the normalized polarizability function
${\tilde \Pi}(x, \eta) = \Pi(q)/D_0$ is given by
\begin{widetext}
\begin{eqnarray}
{\tilde \Pi}(x, \eta) &=&  f(x, \eta) + \theta(x-1) g(x, \eta),  \nonumber \\
f(x, \eta) &=& \left[ 1 - x^2 \eta + x^4 (2 + \eta + 2 \sqrt{1 + \eta}) \right]^{1/2}
 - \ln\left[ \frac{2 \sqrt{1 + \eta x^2}}{-1 + \sqrt{1 + \eta}} \right] - \frac{1}{2} 
 + \frac{3 x^2 \eta -1}{2 x \sqrt{\eta}} \arctan(x \sqrt{\eta}) 
\nonumber \\
&& \mbox{} + \sqrt{1 - x^2 \eta} \left( 2 {\rm arctanh}(\sqrt{1 - \eta x^2}) - {\rm arcsinh}\left[ \frac{\sqrt{1-x^2 \eta}(-1 + \sqrt{1 + \eta})}{x^2 \eta}\right] \right) + \sqrt{1+ \eta}
\nonumber \\
g(x, \eta) &=&  \frac{-\sqrt{x^2 -1} (1 + \eta + 2x^2 \eta - \sqrt{1+\eta})}{2 x (\sqrt{1 + \eta} -1)} + \frac{(3x^2 \eta -1)}{2 x \sqrt{\eta}} \arccos\left[ \sqrt{\frac{1 + \eta}{1 + x^2 \eta}}\right]
 \nonumber \\ && \mbox{}  +
{\rm arctanh} \left[ \frac{x \sqrt{x^2 -1} \eta}{1 + x^2 \eta - \sqrt{1 + \eta}} \right].
\end{eqnarray}
\end{widetext}

\begin{figure}[t]
\includegraphics[width=3in]{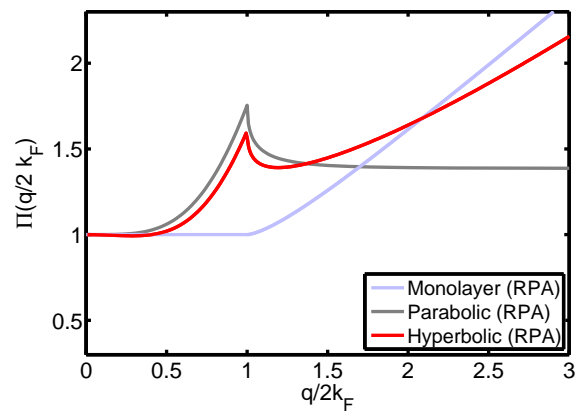}
\caption{\label{Fig:Pi} 
The Random Phase Approximation polarizability function $\Pi(q)$ normalized by the 
density of states for monolayer graphene  and for bilayer graphene with a
hyperbolic dispersion.  Also shown is the
parabolic approximation for the bilayer, which can be obtained from
the hyperbolic dispersion when $\eta \ll 1$.  The
Thomas-Fermi approximation discussed in the text corresponds to the
assumption that the normalized $\Pi(q) = 1$ for all $q$.}
\end{figure}

\begin{figure}[!h]
\includegraphics[width=3in]{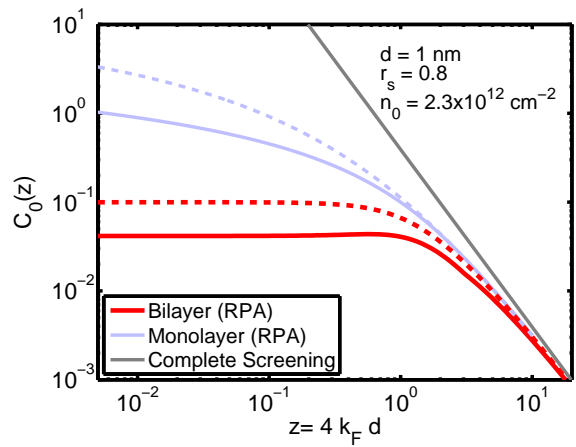}
\caption{\label{Fig:CO} 
Potential autocorrelation function $C[0]$ for monolayer and bilayer graphene.
At the Dirac point, $k_{\rm F}$ is the Fermi wavevector arising from the 
effective carrier density i.e. $k_{\rm F} = \sqrt{\pi n_{\rm eff}}$.  For large density 
$4 k_{\rm F} r_s d \gg 1$, both the monolayer 
and bilayer results approach the ``complete screening'' limit, defined 
here as $C(0) = (4 k_{\rm F} r_s d )^{-2}$.  Notice that the Thomas-Fermi 
approximation shown as dashed lines captures the correct qualitative behavior, 
but can give significantly larger values for $C[0]$, and is therefore unsuitable 
for quantitative comparisons.}
\end{figure}

The polarizability functions for monolayer and bilayer graphene are
shown in Fig.~\ref{Fig:Pi}.  What is left is to calculate the
effective residual density $n_{\rm eff}$ as a function of impurity
concentration.  As discussed earlier, this is obtained by first
calculating the autocorrelation function $C(0)$ from
Eq.~\ref{Eq:main}, using the RPA results shown in Fig.~\ref{Fig:Pi}.
Figure \ref{Fig:CO} shows the autocorrelation function $C(0)$
obtained using the RPA results (solid lines) as well the Thomas-Fermi
results (dashed lines).  We note that except at very high density (where both
monolayer and bilayer graphene approach the ``complete screening''
limit, with $C(0)= (4 k_{\rm F} r_s d )^{-2}$), the Thomas-Fermi
approximation grossly underestimates the effect of screening.  Moreover, 
for typical densities in bilayer graphene $C(0)$ is approximately
constant and independent of carrier density consistent with the 
heuristic picture discussed at the end of Sec.~\ref{Sec:Formalism}.

The effective density calculated within the RPA is shown in Fig.~\ref{Fig:nrms}.  For bilayer 
graphene, we find that the following empirical relationship adequately describes the numerical 
results

\begin{equation}
n_{\rm eff} = \sqrt{ n_{\rm imp}~n_1},
\end{equation}  

\noindent where $n_1 = 11.5 \times 10^{11}~{\rm cm}^{-2}$ for the Thomas-Fermi
approximation, and $n_1 = 6.8 \times 10^{11}~{\rm cm}^{-2}$ for the
RPA results.  The scaling of the bilayer effective density $n_{\rm
  eff} \sim \sqrt{n_{\rm imp}}$ can be anticipated for the TF
approximation in the limit $n_{\rm eff} \ll n_0$.  However, it is
surprising that the simple empirical relation continues to hold both
for the RPA screening theory, and for larger values of $n_{\rm eff}$.

\begin{figure}[!h]
\includegraphics[width=3.2in]{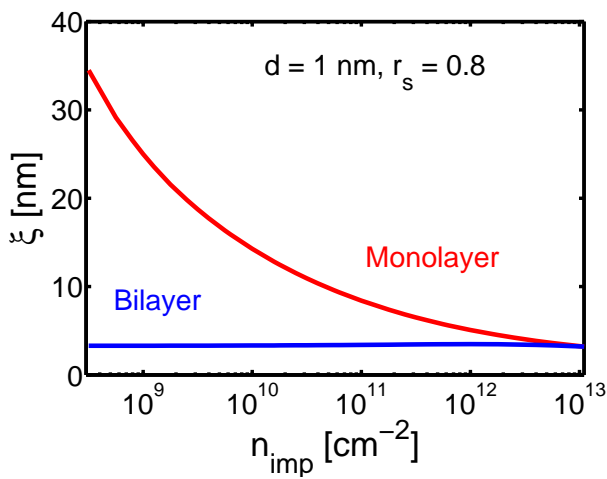}
\caption{\label{Fig:Corr} 
Theoretical results for the puddle correlation length at the Dirac point as a function of impurity 
concentration.  While the puddle size in bilayer graphene ($\xi \approx 3.5~{\rm nm}$) is 
relatively insensitive to the disorder concentration, the size of the puddles in monolayer
graphene varies from $3~{\rm nm}$ in dirty samples to more than $35~{\rm nm}$ in clean 
samples.}
\end{figure}

This dependence of $n_{\rm eff} \sim \sqrt{n_{\rm imp}}$ for bilayer
graphene should be contrasted with similar results obtained previously
for monolayer graphene,\cite{kn:adam2007a} where $n_{\rm eff} = 2
r_s^2 C[0] n_{\rm imp}$ cannot be captured by a similar empirical fit.
For comparison, these earlier results are also shown in Fig.~\ref{Fig:nrms},
where we emphasize that for a given impurity concentration ($n_{\rm
  imp}$), bilayer graphene exhibits larger density fluctuations
($n_{\rm eff}$) than monolayer graphene.  Finally, using
Eq.~\ref{Eq:main}, we can also calculate the puddle correlation
function, and the corresponding HWHM correlation length, $\xi$, that
is shown in Fig.~\ref{Fig:Corr}

\section{Comparison With Experiments}  
\label{Sec:Compare}

\begin{figure}
\includegraphics[width=3in]{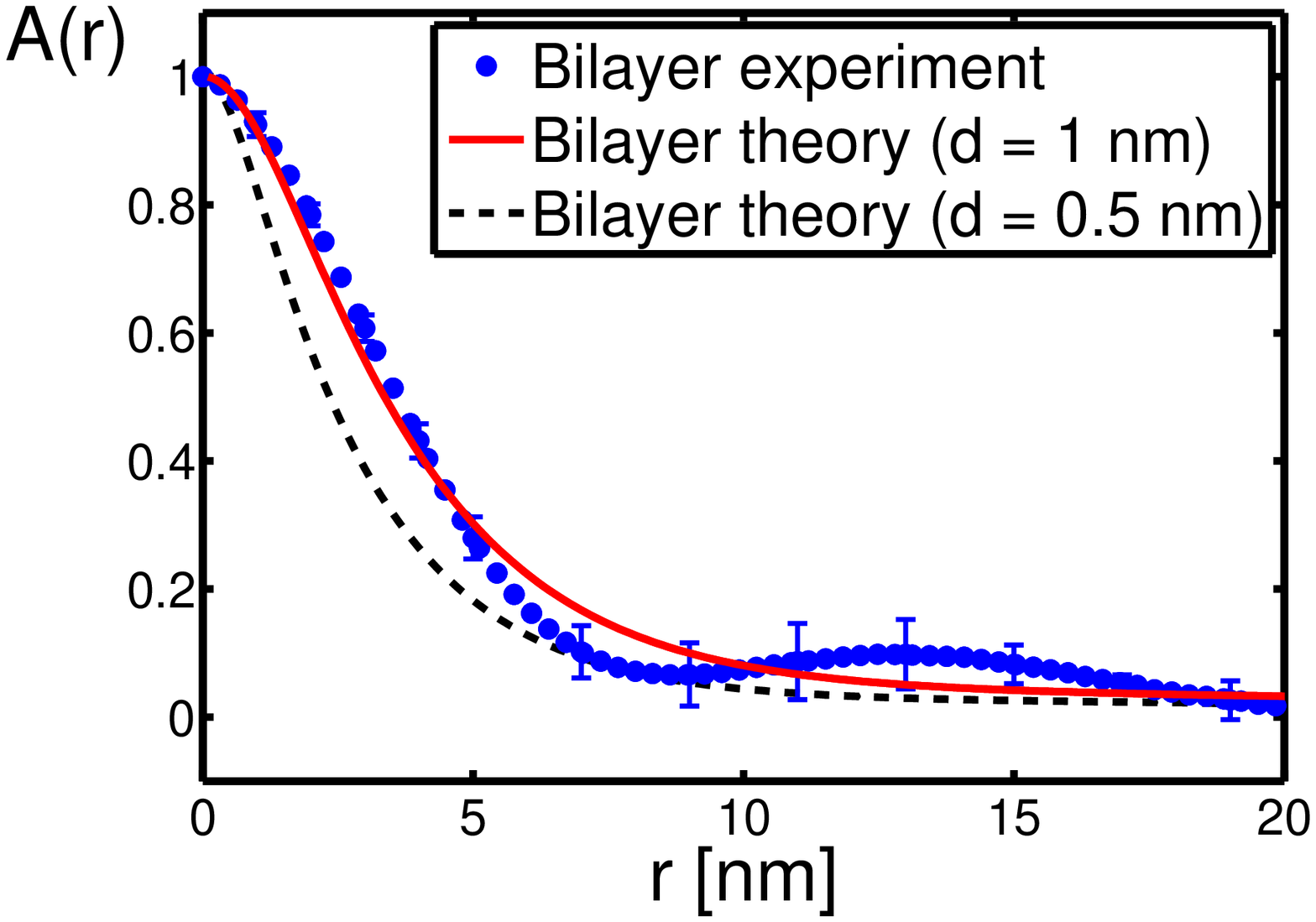}
\includegraphics[width=3in]{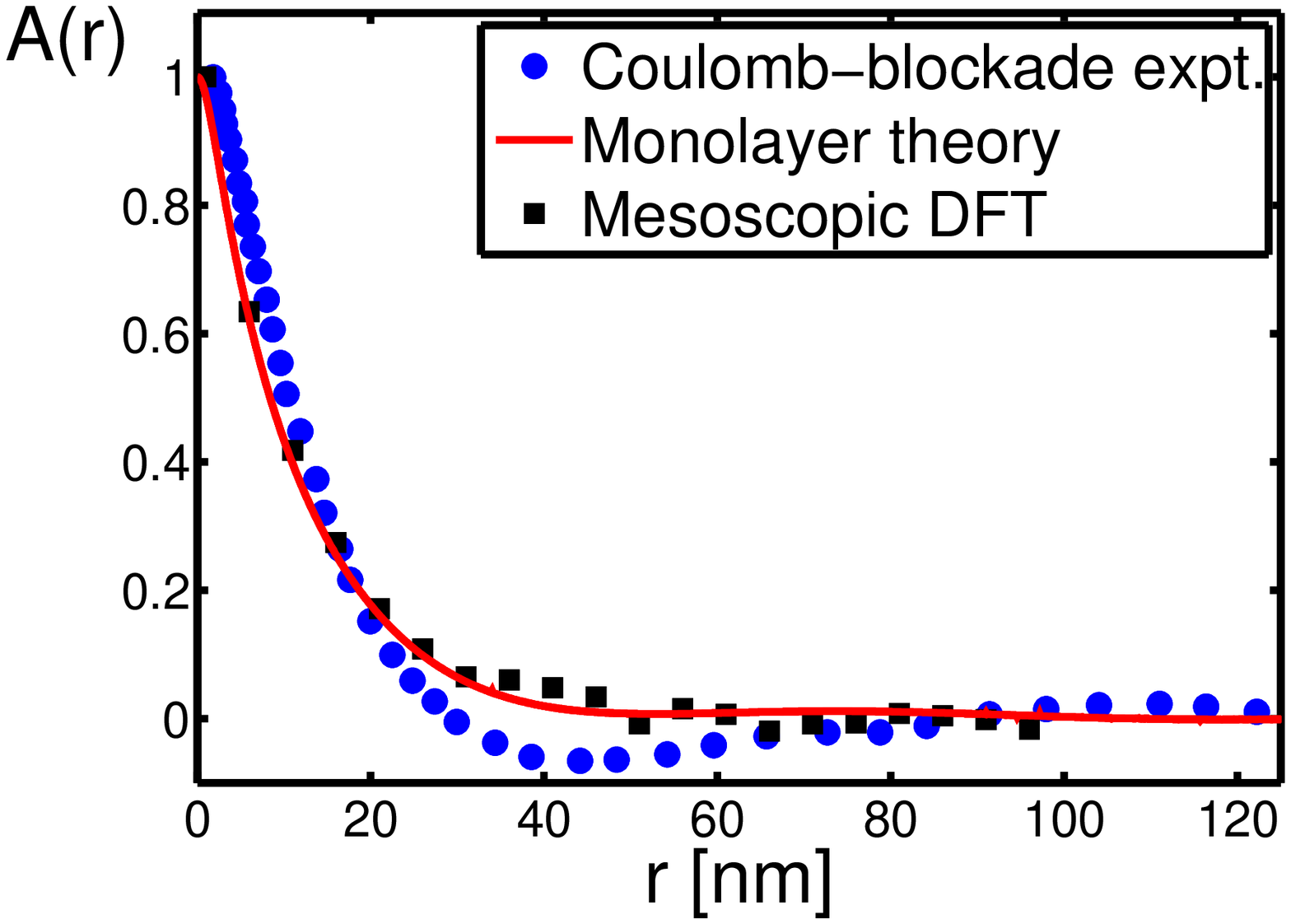}
\caption{\label{Fig:Compare} 
Comparison of theoretical results with experimental data.  Top panel
shows the normalized correlation function $A(r) = C(r)/C(0)$ for
bilayer graphene.  The circles are from the experimental data and solid curve is the
theory for bilayer graphene with $d = 1~{\rm nm}$.  The theory curve
is insensitive to impurity concentration and doping away from the
Dirac point.  The error bars indicate single standard deviation 
uncertainties.\cite{kn:footnote3}  The small oscillation in the data over the monotonic 
decrease is a result of the finite size of the experimental image.  
Bottom panel is the normalized puddle correlation
function in monolayer graphene at the Dirac point.  Note the change in $x$-axis scale
from bilayer graphene in top panel.  The solid curve
is obtained from the self-consistent screening theory.  The black squares are the
results of a numerical mesoscopic density functional theory
calculation for the ground-state properties of monolayer
graphene,\cite{kn:rossi2008} while the circles are experimental
data taken from Deshpande {\it et al.}(Ref.~\onlinecite{kn:deshpande2011}).
Transport measurements on that same device set $n_{\rm imp} = 10^{11}
{\rm cm}^{-2}$ which is the value used for the theory curves.
The theory also uses $d=1~{\rm nm}$, which is the typical 
distance of the impurities from the graphene sheet extracted
from transport measurements of graphene on SiO$_2$.\cite{kn:tan2007}}
\end{figure}

We now compare the calculated correlation functions with experiment.
Figure ~\ref{Fig:Compare}(a) shows that for bilayer graphene, $A(r)$
extracted from the data reported in Ref.~\onlinecite{kn:rutter2011} agrees with the
calculation for $d= 1~{\rm nm}$.\cite{kn:footnote3}  The circles show
the experimental data and the RPA theory for bilayer graphene is shown
for $d = 1~{\rm nm}$ (solid curve) and $d=0.5~{\rm nm}$ (dashed
curve).  The theoretical results are insensitive to the impurity
concentration $n_{\rm imp}$ and to how far the doping is away from the
Dirac point.  Consequently, the only free parameter in the theory is
the distance $d$ of the impurities from the graphene sheet.

In Ref.~\onlinecite{kn:rutter2011}, we reported a maximum peak-to-peak carrier
density fluctuation of $3.6 \times 10^{11}~{\rm cm}^{-2}$.  To extract
an impurity density (using the results in Fig.~\ref{Fig:nrms}), we need to
estimate $n_{\rm rms}$ from this peak-to-peak value.  By assuming that 
the carrier density has a Gaussian distribution, and estimating that the 
peak-to-peak corresponds to a measurement of $4 \sigma$, we roughly 
estimate that $n_{\rm eff} = n_{\rm rms} \approx 10^{11}~{\rm cm}^{-2}$ and
that $n_{\rm imp} = n_{\rm rms}^2/n_1 \approx 1.2 \times 10^{10}~{\rm cm}^{-2}$ for 
the substrate induced impurities in that experiment.  

We can use this value to predict theoretically the corresponding
density fluctuations in the adjacent monolayer sample reported in
Ref.~\onlinecite{kn:jung2010}.  However, one complication is that the
theory discussed in Sec.~\ref{Sec:RPA} was developed for monolayer
graphene at the Dirac point, while the experimental data was taken at
a backgate induced density $n_g = 1.4~\times 10^{12}~{\rm cm}^{-2}$.
Very far from the Dirac point, i.e. when $n_{\rm imp}/n_g \rightarrow
0$, the potential fluctuations $V_{\rm rms}$ can be obtained from
Eq.~\ref{Eq:main} by setting $n_{\rm eff} = n_g$ on the right-hand
side.  In this case, the density fluctuations are
\begin{equation}
n_{\rm rms} = \frac{2 V V_{\rm rms}}{\pi \hbar^2 v_{\rm F}^2}.  
\label{Eq:highdensity}
\end{equation}
\noindent We note that when $z \sim d \sqrt{n_g} \gg 1$, we can use the result 
$C_0(z) = z^{-2}$ (see Fig.~\ref{Fig:CO}) to obtain 
$n_{\rm rms} \approx \sqrt{n_{\rm imp}/(8 \pi d^2)}$.  However, these constraints are not
fully satisfied in the experimental data.  Calculating $n_{\rm rms}$ in the crossover between 
the limits $n_g = 0$  and $n_g \gg n_{\rm imp}$  is more complicated.  For our purposes, it is 
sufficient to extrapolate between the low-density and high-density limits by adding the two 
contributions in quadrature, and solving for $n_{\rm rms}$ self-consistently.  This procedure gives 
\begin{eqnarray}
n_{\rm rms} = 2 r_s \sqrt{n_{\rm imp} C^{\rm RPA}(0)} \left[ 2 n_g + 3
  r_s^2 n_{\rm imp} C^{\rm RPA}(0) \right]^{1/2},
\label{Eq:crossover}
\end{eqnarray}
where the superscript indicates that the RPA screening approximation
has been used. In the limit that $n_g \gg n_{\rm imp}$, Eq.~\ref{Eq:crossover} reduces
to Eq.~\ref{Eq:highdensity}, while in the opposite limit $n_g \rightarrow 0$, 
Eq.~\ref{Eq:crossover} reduces to results shown in Fig.~\ref{Fig:nrms}.

Using the values for $n_{\rm imp}$ and $d$ determined from the bilayer
data discussed above, and using Eq.~\ref{Eq:crossover} for $n_{\rm rms}$ and Eq.~\ref{Eq:main}
to calculate $\xi$, we find theoretically 
(without any adjustable parameter) that $n_{\rm rms} \approx
6~\times 10^{10}~{\rm cm}^{-2}$ and $\xi = 8~{\rm nm}$, which should
be compared to the experimental values extracted from the data
reported in Ref.~\onlinecite{kn:jung2010}.  The area surveyed in
Ref.~\onlinecite{kn:jung2010} was not large enough to obtain $\xi$
accurately.  However, by looking at different real-space cuts of the
autocorrelation function, we conclude that the experimental data is
consistent with a correlation length $6~{\rm nm} < \xi < 11~{\rm nm}$.  This is in
qualitative agreement with our theoretical calculations.  This result
should be contrasted with bilayer graphene shown in
Fig.~\ref{Fig:Compare}(a) where the experimentally determined $\xi =
(3.68 \pm 0.03)~{\rm nm}$, and the $d=1~{\rm nm}$ RPA theory gives
$\xi = 3.4~{\rm nm}$.

To further confirm our results, we compare our calculations to
  measurements made using scanning Coulomb blockade spectroscopy on
a sample of monolayer graphene.\cite{kn:deshpande2011}
  The circles in Fig.~\ref{Fig:Compare}(b) are experimental data for the
  normalized correlation and the solid line is the self-consistent
  theory discussed above using $n_{\rm imp} = 10^{11} {\rm cm}^{-2}$
  and $d=1~{\rm nm}$ for monolayer graphene at the Dirac point.  The
  impurity concentration and impurity distance were determined from
  transport measurements\cite{kn:deshpande2011,kn:adam2007a} and as
  such, {\it no adjustable parameters} were used in the calculation.

In Fig.~\ref{Fig:Compare} we also show (black
squares) the results extracted from a numerical mesoscopic density
functional theory\cite{kn:rossi2008} using the same parameters.  The
agreement between the two calculations provides {\it a posteriori}
justification for our assumption of a global screening function
characterized by the density $n_{\rm eff}$.

\section{Conclusions}
\label{Sec:Conclude}

We conclude with the observation that our results require only that
the source of the disorder potential be uncorrelated charged
impurities, and as such should apply to graphene on other substrates.
For example, recently graphene devices with hexagonal BN gate insulators
have been fabricated showing transport properties similar to suspended
graphene\cite{kn:dean2010} and larger puddles than on SiO$_2$
substrates.\cite{kn:xue2011,kn:decker2011} These observations are
consistent with both a much smaller charged impurity density $n_{\rm
  imp}$ on the BN substrate and with a larger distance $d$ of the
impurities from the graphene layer.  Both these scenarios are possible
because the BN substrate is typically placed on top of the usual
SiO$_2$ wafer which would have similar charged disorder to the samples
we study here.  We argue that an analysis similar to what we have
performed here would be able to uniquely determine both $d$ and
$n_{\rm imp}$.  Moreover, if a similar experiment is done with bilayer
graphene on BN substrates, we predict that the puddle characteristics
will not change much from what we find here with bilayer graphene on
SiO$_2$.

\section*{Acknowledgements}

This work is supported in part by the NIST-CNST/UMD-NanoCenter
Cooperative Agreement.  It is a pleasure to thank W. G. Cullen,
M. S. Fuhrer, and M. Polini for discussions, and P. W. Brouwer,
E. Cockayne, G. Gallatin, J. McClelland and R. McMichael for comments
on the manuscript.


\begin{thebibliography}{26}
\bibitem[{kn:({\natexlab{a}})}]{kn:footnote3}
 The dominant uncertainty in $A(r)$ and $\xi$ comes from
the finite size of the experimental puddle images.  The error in
$A(r)$ is estimated by dividing the image area into $N=4$ patches and
analyzing each patch separately.  Treating the $N$ patches as
independent samples, we compute the variance of the set and estimate
the variance of the whole sample as the variance of the $N$
independent samples, divided by $N-1$.  The uncertainty in $A(r)$ is
square root of the variance, and the uncertainty in $\xi$ then
follows.  We say that the experimentally determined $A(r)$ agrees with
the calculated $A(r)$ for $d = 1~{\rm nm}$ in
Fig.~\protect{\ref{Fig:Compare}}a because theory curve lies within the
experimental uncertainty of $A(r)$.
\end{thebibliography}

\end{document}